# Phonon Dispersion Calculation for Binary Alloys Using WDM Approach

M. Aziziha[1], S. Akbarshahi[1]

[1]Department of Physics & Astronomy, West Virginia University, Morgantown, WV 26506, USA

**Abstract**

The lattice dynamics of AgPd, $Ni_{55}Pd_{45}$, $Ni_{95}Pt_{05}$, and $Cu_{0.715}Pd_{0.285}$ intermetallic have been investigated using the DFT calculation. The phonon dispersions and phonon densities of states along for two symmetry directions are calculated by Weighted Dynamical Matrix (WDM) and compared with virtual crystal approximation (VCA), supercell approach, and inelastic neutron scattering experimental results. The impact of mass, force-constant fluctuation, and Ag concentration on lattice dynamics of AgPd are discussed, and a comparison between WDM and Supercell approach is performed. The averaged first Nearest Neighbor (1NN) force constants between various pairs of atoms in these intermetallic are obtained from the WDM approach. Based on our results, the WDM approach agrees well with the supercell approach, and neutron scattering experimental data. VCA overestimates in some cases and underestimates, in other cases, the first-principles frequencies.

**Introduction**

The study of alloys has drawn a lot of attention, and exotic properties are produced[1–4]. The phonon properties in ordered inter-metallic and disordered alloys calculation and analysis are one of the steppingstones in the materials research. The energy dispersion of phonons provides rich information about the dynamical properties of the material, the ordering behavior, phase stability, and elastic properties which is acritical input in the calculation of thermodynamic properties like the heat capacities, thermal expansion coefficients, transport properties like diffusivity, and quantities like the electron-phonon interaction, etc. To understand the material properties and the distinct phenomena in the microscopic scale, such as their lattice dynamics require robust and accurate computational tools. For ordered crystals, the theory of lattice vibrations has been set up on a rigorous basis. However, for disordered lattice, no comprehensive, computationally efficient approach that can come up with the difficulties of calculations for these systems, is developed, and more research in this field is necessary.

The disordered lattice is commonly seen in alloy systems, which distorts the chemical environment in the lattice and leads to the change in the physical properties. First-principles calculations for such a disordered system require some approximations. A most common approach is the supercell approximation, which makes disordered configurations in a supercell and apply the periodic boundary conditions. The biggest challenge of such calculations is the very large supercells computational load to mimic the distribution of local chemical environments. The size of Supercell strongly depends on the chemical disorder and tends to be computationally very demanding. The approximation approaches are used to overcome this challenge. The Weighted Dynamical Matrix (WDM) [5–7], the virtual crystal approximation (VCA)[8], and the coherent potential approximation (CPA)[9] consider an effective medium for the disordered system. These methods, however, generally do not explicitly consider the local environment around each atom, which is sometimes critical for quantitative evaluations of the physical properties of the disordered system. The VCA is computationally efficient but is not appropriate for non-isoelectronic parent compounds, whereas the supercell approach is accurate but computationally demanding. In general, the VCA is closely tied to the pseudopotential approximation, which means VCA is only reliable if the considered valence orbitals are spatially reasonably similar.

The weighted dynamical matrix approach has advantages over other approximation approaches, which are explained in this work. The most prominent pros are as follows: this approach is straightforward to implement, the generation of a new pseudopotential is not needed for each virtual atom, unlike VCA, and we can calculate any percentage of the alloy without any additional calculation, unlike VCA and Supercell approaches. We compare WDM results to VCA, Supercell, and the experimental reported results for AgPd, $Ni_{55}Pd_{45}$, $Ni_{95}Pt_{05}$, and $Cu_{0.715}Pd_{0.285}$ Alloys.

**Computational Details**

We performed the density functional theory calculations[10,11] with a plane-wave basis set, as implemented in the Quantum-Espresso (QE) code[12]. We employed the Perdew–Burke–Ernzerhof generalized gradient approximation (GGA) exchange-correlation functional[13,14] and Optimized Norm-Conserving Vanderbilt Pseudopotential (ONCVPSP)[15]. A variable cell-structure relaxation was performed in QE until the Hellmann-Feynman force and stress are less than 1mRy/Bohr and 0.1 mRy/Bohr[16,17]. The relaxed primitive unit cell with cubic structure ($Fm\bar{3}m$) (225) of Ni, Pd, Pt, Cu, Ag is used to construct a 4×4×4 supercell containing 64 atoms. Also, Spin-polarization considered. The relaxation of this Supercell is done with a $6 \times 6 \times 6$ Monkhorst-Pack $k$-point grid[18]. The energy cut-off 110 Ry for wave functions was employed for calculations. To evaluate the force constants using PHONOPY [19] software. In the supercell method for alloys, a finite-size supercell with defects breaks the space group symmetry and leads to a shrinking BZ in reciprocal space. The state-of-the-art band unfolding methods have been developed for electronic problems to recover the phonon spectra within the BZ of the primitive cell[20,21] as well as for phonon problems[22,23]. Here, we use the unfolding program developed by Ikeda et al. to carry out the phonon band unfolding. For the supercell calculation of the AgPd, we used from Sai Mu et al. data[1].

**Weighted Dynamical Matrix (WDM) Approach**

To find the phonon modes, one needs to construct the dynamical matrix. The dynamical matrix $\mathbf{D}(\mathbf{q})$ at the wavevector $\mathbf{q}$ is constructed as follows:

$$D_{ii'}^{\alpha\beta}(\mathbf{q}) = \frac{1}{\sqrt{m_i m_{i'}}} \sum_{l'} \Phi_{\alpha\beta}(0i, l'i') \exp[i\mathbf{q} \cdot (\mathbf{r}_{l'i'} - \mathbf{r}_{0i})], \quad (1)$$

where $m_i$ is the mass of the $i^{th}$ atom. Phonon frequencies $\omega(\mathbf{q}, \kappa)$ and mode eigenvectors $\boldsymbol{\chi}(\mathbf{q}, \kappa)$ at $\mathbf{q}$, where $\kappa$ is the band index, are obtained by solving the eigenvalue equation:

$$\mathbf{D}(\mathbf{q}) \boldsymbol{\chi}(\mathbf{q}, \kappa) = [\omega(\mathbf{q}, \kappa)]^2 \boldsymbol{\chi}(\mathbf{q}, \kappa). \quad (2)$$

To calculate the phonon modes for these alloy samples, first, the Hellmann-Feynman forces of the parent structures. The weighted dynamical matrix is constructed as follows:

$$\overline{D}_{ii'}^{\alpha\beta}(\mathbf{q}) = \frac{1}{\sqrt{m_i m_{i'}}} \sum_{l'} \overline{\Phi}_{\alpha\beta}(0i, l'i') \exp[i\mathbf{q} \cdot (\mathbf{r}_{l'i'} - \mathbf{r}_{0i})], \quad (3)$$

**Results and Discussion**

For $Ni_{55}Pd_{45}$, $Ni_{95}Pt_{05}$, and $Cu_{0.715}Pd_{0.285}$, we compare our proposed weighted dynamical matrix (WDM) results with the virtual crystal approximation (VCA) and with inelastic neutron scattering experimental results[24–26]. For AgPd, we compare the results of WDM with Supercell computed results, which are performed and reported by Sai Mu et al.[1].

The results obtained for the phonon dispersion curves and phonon density of states along [0,0,$\zeta$] and [0,$\zeta$,$\zeta$] directions are shown in Fig. 1(a-d). The WDM computed, and the experimental results of phonon frequencies agree reasonably well both along [0,0,$\zeta$] and [0,$\zeta$,$\zeta$] directions ($\zeta = \frac{\vec{q}}{\vec{q}_{max}}$, $\vec{q}$ is the phonon wavevector). However, VCA calculations show an overestimation of the NiPd and CuPd and an underestimation for NiPt. Two peaks in the FCC structure indicate the vibrational resonance due to atomic motion. The highest frequency peak in the DOS plot corresponds to the Longitudinal mode where the phonon band is flat, and the lower frequency peak is related to the top of the lower-lying Transverse modes. Fig. 3 shows the impact of Ag concentration on the phonon dispersion of AgPd along $[0, 0, \zeta]$ and $[0, \zeta, \zeta]$ directions, calculated by the WDM approach. With increasing the Ag concentration, a decrease in the phonon frequency is observed.

The results obtained from the phonon dispersion curves along $[0,0,\zeta]$ and $[0,\zeta,\zeta]$ directions by WDM and Supercell approaches calculations are shown in Fig. 2(a-d). The plotted phonon dispersions confirm that both the effect of force constant and mass fluctuations are insignificant.

The first nearest-neighbor force constants, which shown in Table 1, are an order of magnitude larger than those of the further neighbors. Hence, we used the weighted average of the force constant. We use the first nearest-neighbor force constants and ignore the further neighbors. Then, we compare the force constants and conclude that the Ni-Pd force constants (-0.718) should be less than the Ni-Ni (-0.734) ones. That is because Pd is larger than Ni, and the Ni-Pd bonds are larger than the Ni-Ni bonds.

**Conclusions**

We have investigated the lattice dynamics AgPd, $Ni_{55}Pd_{45}$, $Ni_{95}Pt_{05,}$ and $Cu_{0.715}Pd_{0.285}$ inter-metallics using first-principles density functional theory. The agreement between WDM, supercell approach, and neutron scattering experiments is good. VCA overestimate and underestimate first-principles frequencies. It is because of the mixing pseudopotential before calculation. This is important from the point of view of the feasibility of using *ab initio* ordered alloy force constants to study the disordered AgPd and impact of the Ag concentration on the phonon dispersion AgPd. Also, we can see the mass and force-constant fluctuation does not have a significant effect on the phonon dispersion. The results have been analyzed from the nearest-neighbor force constant values between various pairs of species. The behavior of force constant versus bond distance follows the expected trend.

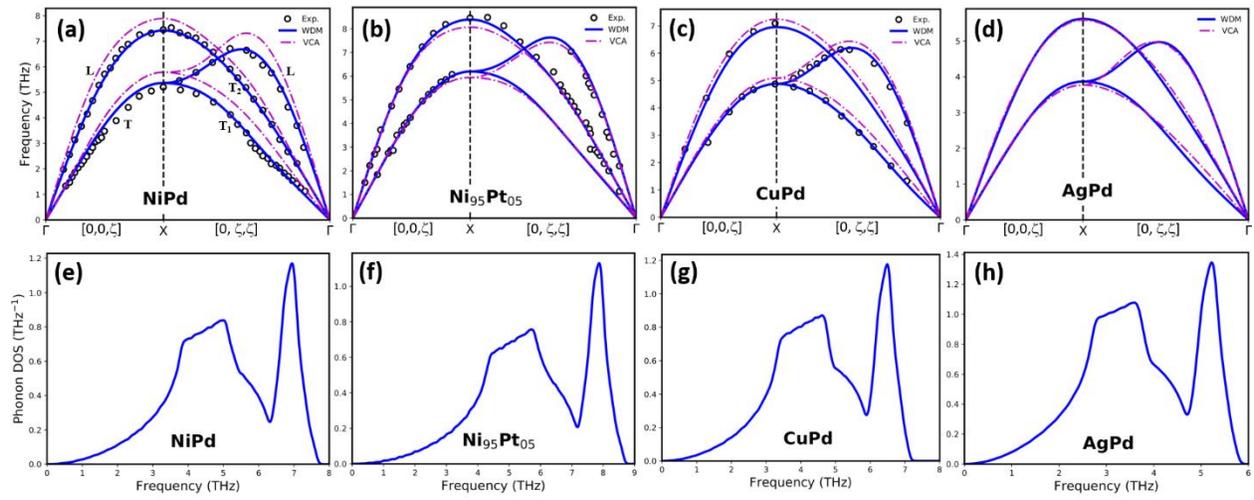

Fig. 1 (a)-(d) Comparison of phonon dispersion of the $Ni_{55}Pd_{45}$, $Ni_{95}Pt_{05}$, CuPd and AgPd along $[0, 0, \zeta]$ and $[0, \zeta, \zeta]$ directions between WDM and VCA approach. The blue line in the figures represent the WDM calculations, the magenta dashed-dotted line in the figures represent the VCA calculations and the circle in the figures represent the experimental results. (e)-(h) Phonon density of state NiPd, $Ni_{95}Pt_{05}$, CuPd and AgPd.

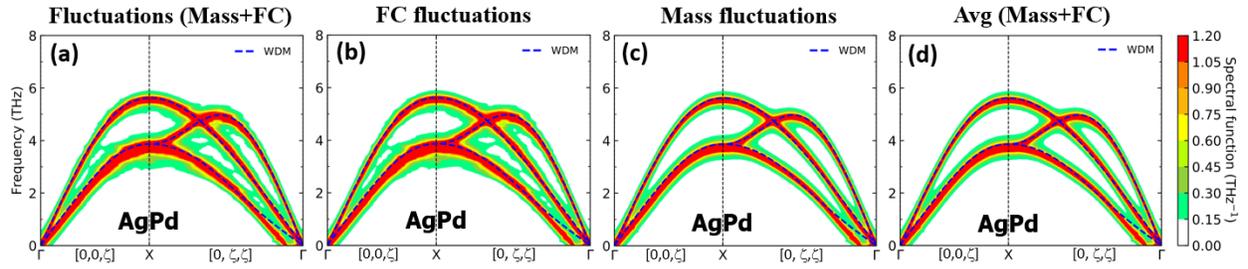

Fig. 2 Comparison of phonon dispersion of AgPd along [0, 0, ζ] and [0, ζ, ζ] directions between Supercell and WDM approach and Impact of mass fluctuations and force constant fluctuations on the phonon dispersion. The blue dashed line in the figures represent the WDM calculations. (a) phonon dispersion including both mass and force constant fluctuations, (b) phonon dispersion including only force constant fluctuations, (c) phonon dispersion including only mass fluctuations, and phonon dispersion including averaged mass and force constants.

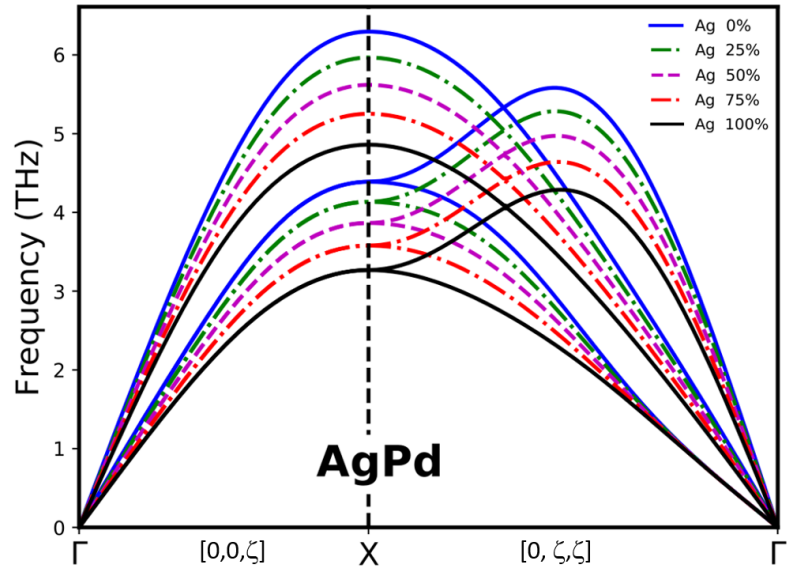

Fig. 3 Impact of Ag concentration on the phonon dispersion of AgPd along $[0, 0, \zeta]$ and $[0, \zeta, \zeta]$ directions calculated by WDM approach. The blue, green, magenta, red and black line represent the 0, 25, 50, 75, 100 percent of the Ag respectively.

Table 1 The averaged first Nearest Neighbor (1NN) force constants in Pt, Ni, Pd, Cu, Ag extracted from supercell calculations. The force constants are given in eV/Å$^2$.

| Nearest Neighbor force constants | Average | Pt-Pt | Ni-Ni | Pd-Pd | Cu-Cu | Ag-Ag |
|---|---|---|---|---|---|---|
| 1NN | $\Phi_{xx,yy,zz}$ | -0.913 | -0.734 | -0.699 | -0.546 | -0.385 |